\documentclass[12pt,preprint]{aastex}
\makeatletter

\newcommand{\Rmnum}[1]{\expandafter\@slowromancap\romannumeral#1@}
\makeatother
\usepackage{amsmath,amssymb}
\slugcomment{Accepted by ApJ} 
\shorttitle{3D simulations of flux rope formation}
\shortauthors{Xia et al.}
\begin{document}
\title{3D PROMINENCE-HOSTING MAGNETIC CONFIGURATIONS: CREATING A HELICAL
  MAGNETIC FLUX ROPE}
\author{C. Xia\altaffilmark{1}, R. Keppens\altaffilmark{1}, Y. Guo\altaffilmark{2}}
\altaffiltext{1}{Centre for mathematical Plasma Astrophysics, Department of 
Mathematics, KU Leuven, Celestijnenlaan 200B, 3001 Leuven, Belgium}
\altaffiltext{2}{School of Astronomy and Space Science, Nanjing University, 
Nanjing 210093, China}

\begin{abstract}
The magnetic configuration hosting prominences and their surrounding coronal
structure is a key research topic in solar physics. Recent theoretical and
observational studies strongly suggest that a helical magnetic flux rope is an 
essential ingredient to fulfill most of the theoretical and observational 
requirements for hosting prominences. To understand flux rope formation details 
and obtain magnetic configurations suitable for future prominence formation 
studies, we here report on three-dimensional isothermal magnetohydrodynamic 
simulations including finite gas pressure and gravity. Starting from a 
magnetohydrostatic corona with a linear force-free bipolar magnetic field, we 
follow its evolution when introducing vortex flows around the main polarities 
and converging flows towards the polarity inversion line near the bottom of the 
corona. The converging flows bring feet of different loops together at the 
polarity inversion line and magnetic reconnection and flux cancellation happens.
Inflow and outflow signatures of the magnetic reconnection process are 
identified, and the thereby newly formed helical loops wind around pre-existing
 ones so that a complete flux rope grows and ascends. When a macroscopic flux 
rope is formed, we switch off the driving flows and find that the system 
relaxes to a stable state containing a helical magnetic flux rope embedded in 
an overlying arcade structure. A major part of the formed flux rope is threaded 
by dipped field lines which can stably support prominence matter, while the 
total mass of the flux rope is in the order of 4--5$\times 10^{14}$ g.
\end{abstract}


\section{INTRODUCTION}\label{intro}

Observations show that prominences are always residing in the lower regions 
of coronal cavities, which are tunnel-like regions of reduced electron density 
along so-called filament channels. When observed at the solar limb, coronal 
cavities appear as relatively dark, elliptical regions around 
prominences underneath coronal streamers \citep{Gibson2010ApJ,
Waldmeier1970SoPh}. Observations on quiescent prominences suggest that 
prominences and their surrounding coronal cavities are hosted by a 
single large-scale structure in the solar corona: a helical magnetic flux rope
\citep{Berger2012ASPC,WangYM2010ApJ}. The latest observations of coronal
magnetometry on cavities found twist or shear of magnetic field extending up 
into the cavity and a pattern of concentric rings in line-of-sight velocity
which may be explained by a magnetic flux rope model \citep{Bak2013ApJ}. In situ
 measurements of interplanetary CMEs showed strong evidence for magnetic flux 
rope structures \citep{Burlaga1982GeoRL}, which means that either there is a 
pre-existing flux rope in a prominence before eruption, or it changes into a 
flux rope structure during its eruption and propagation. Indications from 
aligned chromospheric fibrils in filament channels \citep{Foukal1971SoPh} and 
from direct measurements of prominence magnetic fields \citep{Leroy1983SoPh} 
show that the magnetic field in a filament channel has a strong component along 
the direction of the polarity inversion line (PIL), which is consistent with a
weakly twisted flux rope model. Magnetohydrostatic solutions for the 
support of prominence matter within a 2.5D flux rope have been presented in 
analytical \citep{Low2004ApJ} and numerical \citep{Blok2011AA,Hillier2013ApJ} 
models. Therefore, a helical magnetic flux rope is a promising magnetic 
structure to host a prominence. 

The origin of a magnetic flux rope in the corona has been explained by two
theories: (1) a pre-existing flux rope in the convection zone emerges
through the solar surface into the corona near active regions \citep{Okam08,
Fan01,Fan2010ApJ}; (2) magnetic reconnection and flux cancellation at the PIL
of a sheared magnetic arcade leads to the formation of helical field lines
\citep{vanBa1989ApJ,Marte01}. From observations, \citet{Macka08} found that
most filaments are outside of active regions and their formation is not 
directly related to flux emergence but results from flux cancellation or
coronal reconnection. Observations indicate that flux cancellation, 
which is the mutual disappearance of positive and negative magnetic flux when 
they encounter each other at PILs in photospheric magnetograms, constantly 
happens during the formation of a filament \citep{Wang2007ApJ,Chae2001ApJ}.
Theoretical considerations \citep{vanBa1990ApJ} and 2.5D numerical 
studies creating force-free magnetic field configurations for flux ropes 
embedded in arcades \citep{vanBa1989ApJ} suggested that photospheric mass 
flows may cause restructuring of coronal magnetic fields, leading to flux rope 
formation and eruption. Two kinds of photospheric flows have been 
considered as the main driver: shear flow and converging flow. Shear flow
caused by flux emergence, differential rotation, meridional flow, and magnetic 
diffusion stretches coronal field arcades creating an axial component of the 
magnetic field. Net converging flows toward the PIL in weak field regions is the
result of magnetic elements diffusion caused by random supergranular motions, 
which tend to transport radial magnetic flux from regions of high flux density 
to regions of low flux density \citep{Zirker1997SoPh,Leighton1964ApJ}. 
Converging flows bring opposite-polarity magnetic elements, which are feet of 
different magnetic loops, to collide at the PIL. This drives photospheric 
magnetic reconnection \citep{Litvinenko2007ApJ} (causing the flux cancellation 
phenomenon) to effectively change the topology of magnetic field from arcades 
into helical flux ropes and increases the axial component magnetic field. 

Based on the flux cancellation mechanism, many numerical models 
\citep[e.g.][]{Amari1999ApJ,Amari2000,Amari03a,Amari03b,Amari10,Mackay2006,
Yeates09} have produced twisted flux ropes. They use different driving means 
applied as bottom boundary condition, such as (1) imposing a time-dependent 
decrease of the magnetic flux via prescribed electric fields; (2) imposing 
time-dependent vertical magnetic fields derived from large-scale photospheric 
motions such as differential rotation, meridional flow, and surface diffusion;
 and (3) imposing shearing flows or converging flows at the PIL via prescribed
 velocity fields. Most of these models were solving simplified 
magnetohydrodynamic (MHD) equations. In particular, nonlinear force-free 
(magnetic field only) configurations were used in \citet{Mackay2006,Yeates09} 
to adress the evolution of large-scale to global coronal field solutions due 
to photospheric flux changes. A frequently made assumption is to work in 
zero-$\beta$ conditions, excluding gas pressure and neglecting solar gravity 
\citep{Amari1999ApJ,Amari2000,Amari03a,Amari03b,Amari10,Lione02}, while a 
(topologically oriented) comparison between zero-$\beta$ MHD and nonlinear 
force-free models for a sigmoidal flux rope is found in \citep{Savcheva2012}.
However, the overall equilibrium balance, plasma thermal structure, and 
small-scale dynamics of quiescent prominences may be significantly affected 
by gas pressure and gravity. Selected global models \citep{Linker01,
Linker03} did include gravity and finite beta effects, using typically 
polytropic MHD assumptions (with polytropic indices mimicking near isothermal 
conditions) in 3D, while full thermodynamic effects have only entered reduced 
dimensionality simulations \citep{Linker01,Lione02}. These works usually 
focus on (loss of) global equilibrium and the eruption of prominences, while 
we will concentrate on a local, 3D isothermal model addressing details of 
flux rope formation with a stable flux rope endstate suitable for prominence 
support. Some ideal MHD simulations at finite beta but without gravity 
included produced twisted field lines from arched loops by coronal 
reconnection driven by shearing flows or converging flows at the bottom 
\citep{DeVor00,DeVor05,Welsch05}. But the final assembly of those twisted 
field lines did not resemble typical flux ropes with elliptical 
cross-sections, an aspect which we will demonstrate here, while including 
gravity.

In this paper, we  perform 3D isothermal MHD numerical simulations, at realistic
finite-$\beta$ regimes in a gravitationally stratified solar coronal atmosphere.
Our aim is to generate a helical flux rope self-consistently starting from a 
simple arcade configuration by imposing systematic photospheric motions. The 
first step is to build a sheared magnetic arcade field, using extrapolation 
techniques, augmented with a stratified atmosphere. The second step will 
introduce converging photospheric flows that drive magnetic reconnection along 
the PIL. This is here achieved in finite-$\beta$ MHD, realizing an endstate that
 relaxes to a stable flux rope which can host a prominence.
This paper is structured as follows. In Section \ref{numer} we describe the 
numerical method and the modeling procedure. The results and analysis are
gathered in Section~\ref{resu}. A general description of the flux rope 
formation process is given in Subsection~\ref{frbirth}. Subsection~\ref{prec} 
reports the analysis of the magnetic reconnection that gives birth to the 
large-scale helical flux tubes. Force analysis and twist property are given in
Subsection~\ref{force} and Subsection~\ref{twist} respectively. Conclusions 
and an outlook are given in section~\ref{conclusion}.

\section{NUMERICAL METHOD}\label{numer}

\subsection{Initial setup and governing equations}
The computational domain is a 3D finite box of sizes $-120<x<120$ Mm, 
$-90<y<90$ Mm, and $3<z<123$ Mm in Cartesian coordinates, intended to be large 
enough to cover the coronal part of a complete filament channel. We first set 
up an initial magnetic field by linear force-free field extrapolation from an 
analytic, bipolar magnetogram $B_m=B_z(x,y,z=0,t=0)$ with the magnetic 
field strength range in $\pm20$ G at the height of $z=0$ Mm. $B_m$ is composed 
of two elliptic Gaussian distributions placed across the $x$-axis (the PIL is 
in the $x$-direction): 
\begin{equation}
B_m=B_0 \exp(-\frac{(x-x_1)^2}{2 \delta x^2}-\frac{(y-y_1)^2}{2 \delta 
y^2})-B_0 \exp(-\frac{(x-x_2)^2}{2 \delta x^2}-\frac{(y-y_2)^2}{2 \delta y^2}),
\end{equation}
where $B_0=20$ G, $x_1=x_2=0$ Mm, $y_1=-y_2=40$ Mm, $\delta x=50$ Mm, and
 $\delta y=20$ Mm.
We use this magnetogram prescription to construct a first 3D linear force-free 
field in the entire box. The linear force-free field is integrated
by the exact Green's function method \citep{Chiu1977ApJ} with constant
$\alpha=-0.08$. Because the pressure scale height in the lower corona is 
relatively large, we simplify the thermodynamics to an isothermal assumption, 
taking the uniform temperature fixed at $T_0=1$ MK. We then still need to 
prescribe a pressure-density variation throughout the box, but from hydrostatic 
equilibrium under gravitational stratification, the density distribution is 
derived to be 
\begin{equation} 
\rho=\rho_b \exp[-\frac{g_\odot r_\odot^2}{R T_0} 
(\frac{1}{r_\odot}-\frac{1}{r_\odot+z})]\,
\end{equation} 
with $\rho_b=2.34\times10^{-15}$ g cm$^{-3}$ the bottom density at $z=0$, 
$g_\odot$ the solar surface gravitational acceleration, $r_\odot$ the solar 
radius, and gas constant $R$. The gas pressure is finite and follows the ideal
gas law $p=\rho R T_0$. There is no flow prescribed initially. This initial 
magnetic configuration, along with information on the density structuring is 
shown in Figure~\ref{finit}(a).

In all subsequent steps, we proceed by solving the isothermal MHD equations 
given by
\begin{align}
 \frac{\partial \rho}{\partial t}+\nabla\cdot\left(\rho\mathbf{v}\right)
       &=0,\\
 \frac{\partial \left(\rho\mathbf{v}\right)}{\partial t}+\nabla\cdot\left(
  \rho\mathbf{vv}+p_{\rm tot}\mathbf{I}-\frac{\mathbf{BB}}{\mu_0}\right)&=
  \rho\mathbf{g}-\nabla\cdot\Pi,\\
 \frac{\partial \mathbf{B}}{\partial t}+\nabla\cdot\left(\mathbf{vB}-
  \mathbf{Bv}\right)&=0,
\end{align}
where $\rho$, $\mathbf{v}$, $\mathbf{B}$, and $\mathbf {I}$ are the plasma 
density, velocity, magnetic field, and unit tensor, respectively, while the 
total pressure is $p_{\rm tot}\equiv p+\frac{B^2}{2 \mu_0}$ and $\mathbf{g}=-
g_\odot r_\odot^2/(r_\odot+z)^2\mathbf{\hat{z}}$ is the gravitational 
acceleration with $r_\odot$ the solar radius and $g_\odot$ the solar surface 
gravitational acceleration. The mass conservation equation and ideal MHD 
induction equation are augmented with a momentum equation where we 
incorporate a stress tensor $\Pi$ having components 
$\Pi_{ij}=-\mu\left(\frac{\partial 
v_i}{\partial x_j}+\frac{\partial v_j}{\partial x_i}-\frac{2}{3}\delta_{ij}
\nabla\cdot\mathbf{v}\right)$ with the dynamic viscosity coefficient $\mu=0.5$ 
g cm$^{-1}$ s$^{-1}$ and the Kronecker delta $\delta_{ij}$. For the 
normalization of the equations, we set the unit of length, number density, 
velocity, and magnetic field to be 10 Mm, $10^9$ cm$^{-3}$, 116.45 km/s, and 2 
G respectively. We use the
parallelized Adaptive Mesh Refinement (AMR) Versatile Advection Code 
\citep[MPI-AMRVAC;][]{Kepp12} to numerically solve these equations. We choose a 
third-order accurate, shock-capturing scheme combining an HLL(C) scheme
\citep{Mignone2006MNRAS} and a third order Cada-limited reconstruction, with a 
three-step Runge-Kutta time marching~\citep{Cada09JCoPh}. The constraint
$\nabla\cdot \mathbf{B}=0$ is controlled by a diffusive approach, which 
diffuses away any numerically generated divergence of magnetic field at the 
maximal rate allowed by the CFL condition \citep{vanderH07, Keppens03}. In 
practice, the magnetic field then gets updated according to $\mathbf{B}+C_d
\left(1/\Delta x^2+1/\Delta y^2+1/\Delta z^2\right)^{-1}\nabla \nabla\cdot 
\mathbf{B}$ where we take $C_d$ of order unity while using centered 
differencing approximations for the gradients. The AMR grid we use has three 
levels and an effective resolution of $512\times384\times256$ with spatial 
resolving ability of 469 km. The automated refining (or coarsening) decisions 
are made upon the local error estimation of magnetic field and density with 
90\% and 10\% weights respectively. The Level 1, Level 2, and Level 3 grids 
cover 74\%, 13.4\%, and 12.6\% of the total volume of the box respectively.

\subsection{Twisting the arcade system}

\begin{figure}
\includegraphics[width=5.8in]{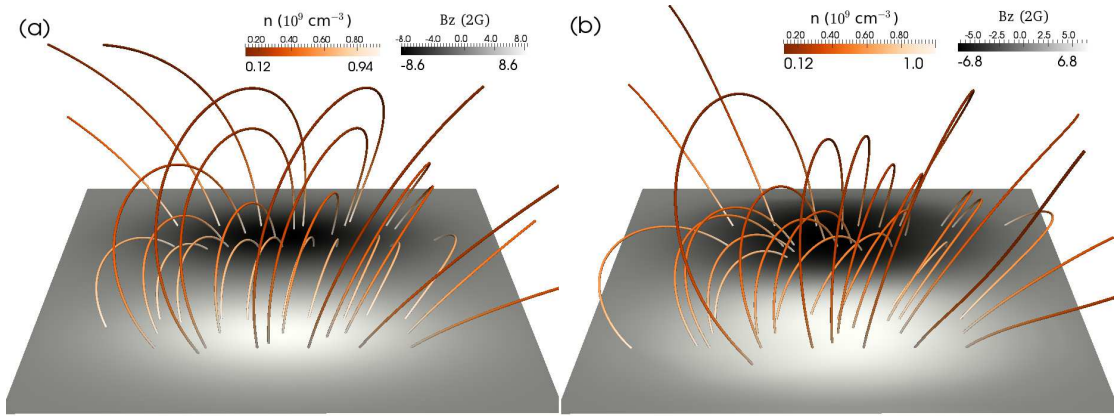}
\caption{(a) Initial linear force-free field as a sheared arcade extrapolated 
from the bottom magnetogram $B_z$ in a gray scale; (b) more realistic arcade 
after twisting motion driven at the bottom. Magnetic field lines in orange and 
white are colored by the number density of plasma. The color bars are scaled to 
the instantaneous data range.
(A color version of this figure is available in the online journal.)
}
\label{finit}
\end{figure}

\begin{figure}
\includegraphics[width=5.8in]{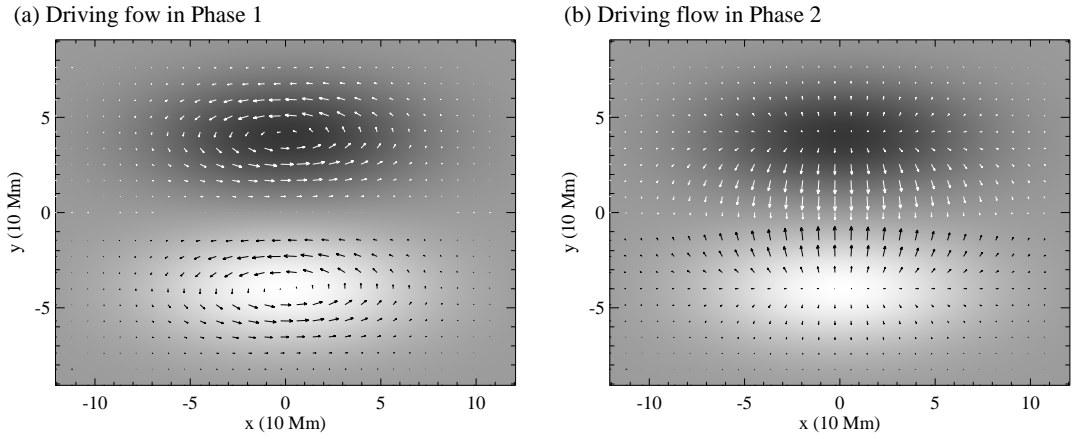}
\caption{(a) twisting and (b) converging velocity field (shown by arrows) on top
 of the magnetogram in a gray scale on the bottom boundary in Phase 1 and 
Phase 2 respectively.}
\label{fdriv}
\end{figure}

\begin{figure}
\includegraphics[width=\textwidth]{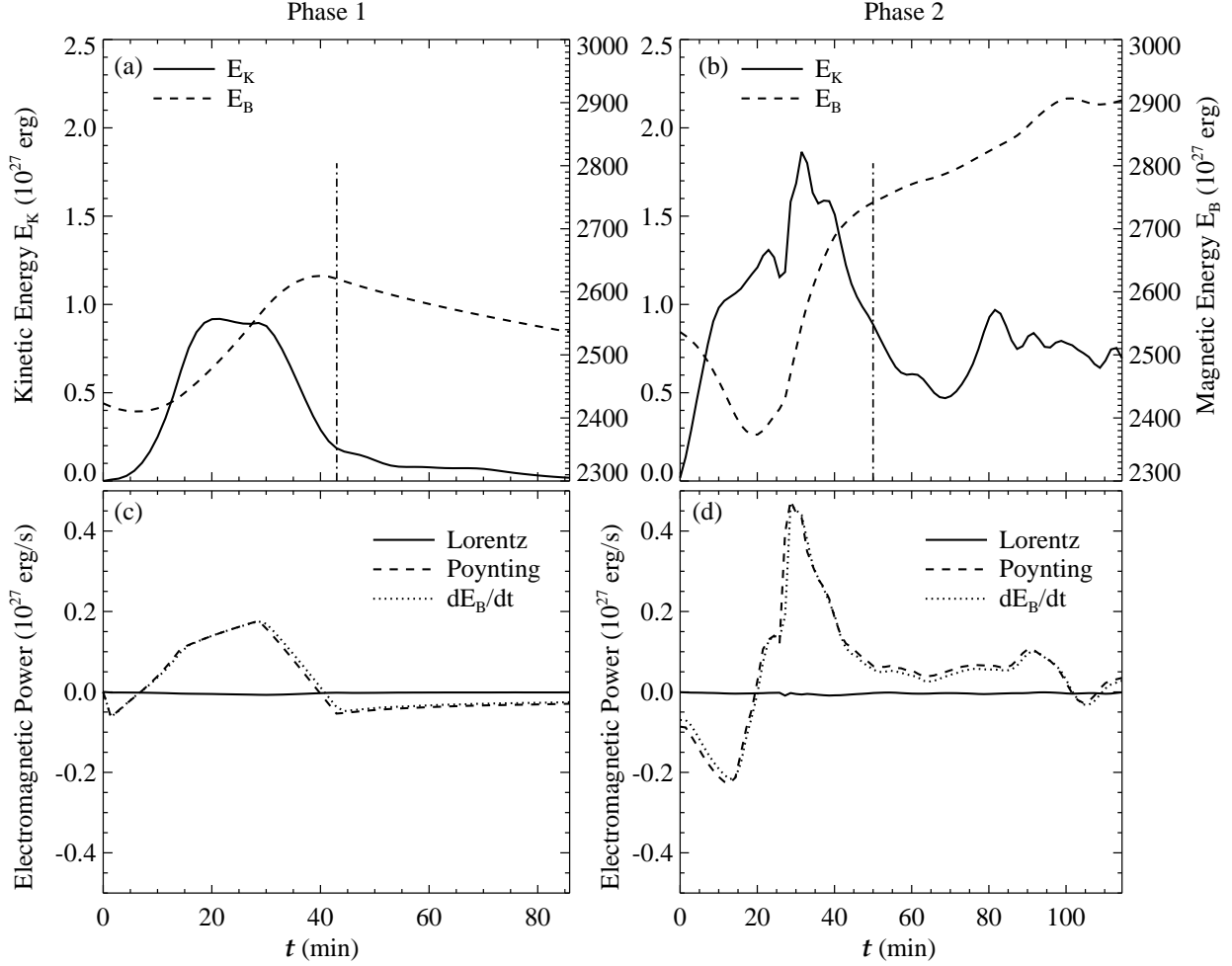}
\caption{(a) Time evolutions of the kinetic energy $E_K$ (solid line) and 
the magnetic energy $E_B$ (dashed line) in the whole domain during Phase 1; (b) 
the similar quantification as in (a) during Phase 2; (c) The Lorentz force power
in reverse sign (solid line), the Poynting flux in reverse sign (dashed line),
and the time derivative of $E_B$ (dotted line) versus time during Phase 1; (d)
a similar quantification as in (c) during Phase 2. The vertical dotted dashed 
lines show the times when bottom driving flows completely stop in each phase. 
The start of Phase 2 is the same as the end of Phase 1.}
\label{fenergy}
\end{figure}

By construction, the initial magnetic arcade is non-potential but linear 
force-free and the topology is such that the larger magnetic loops connecting 
the outer regions in arch-shaped field lines have larger shearing angles 
relative to the PIL than the smaller loops connecting opposite polarities near 
the central PIL. This can be seen in the field lines as shown in Figure
\ref{finit}(a). However observations typically show that the larger/higher 
magnetic loops are more close to potential fields, and in particular exhibit 
smaller shearing angles than the underlying smaller loops \citep{Marti98}. To 
adjust the arcade to this more realistic topology, we impose a horizontal 
twisting velocity field on the bottom boundary, which is composed of two 
large-scale in-plane vortices rotating in the same direction around the two main
 polarities (see Figure~\ref{fdriv}(a)), such that the vertical component of
 the magnetic field at the bottom is preserved \citep{Amari96}. This
twisting flow is not intended to model real flows on the sun. The twisting 
velocity field is formulated as
\begin{equation} \label{eq:6}
 v_x^a=f(t)\frac{\partial \phi}{\partial y},~~~~
 v_y^a=-f(t)\frac{\partial \phi}{\partial x},~~~~
 \phi=C_0 B_m^2 \exp(\frac{B_m^2-B_{max}^2}{\delta B^2}),
\end{equation}
where $B_{max}\approx20 $G is the maximum value of $B_m$, $\delta B=5 B_{max}$, 
$f(t)$ is a linear ramp function with values between 0 and 1 to switch the 
driving smoothly on and off with a ramp of 14.3 min, and the amplitude factor 
$C_0$ is chosen so that the imposed velocity magnitude has a maximum value of 
11.8 km s$^{-1}$ and the maximum initial Alfv\'{e}n Mach number is 0.0126. 
Note that the driving velocity prescription is used to fill the cell-centered 
velocities in the ghost cells of the bottom boundary. In doing so, the velocity 
on the bottom plane, which corresponds to a cell face, can still adopt some 
intermediate value between the first physical 
cell-center value and the imposed ghost cell-center value. This local bottom 
plane value is computed during the (nonlinear) cell-center to cell-edge 
limited reconstruction procedure and next applied in the flux quantifications. 
This means that fluxes at the bottom boundary plane may not be zero when the 
driving flow is switched off, as we will see later. 
We then time advance the isothermal MHD equations from above, augmented 
with further boundary prescriptions. We impose zero vertical velocity on 
the bottom face and zero velocity on the other five faces of the box during 
this phase by antisymmetric boundary condition. For the boundary condition of
the magnetic field, we do extrapolation keeping the normal gradient to be
zero. The scheme is one-sided third-order accurate finite difference
for the bottom and second order for the other faces. Then we modify the 
normal component to fulfill the divergence-free condition discretely using a 
second-order centered difference evaluation. We use zeroth order extrapolation 
for the density on side boundaries, fixed gravitationally stratified density at 
the bottom, and adopt a gravitationally stratified density profile at the top 
extrapolated by second order centered difference. The twisting driving velocity 
imposed at the bottom boundary lasts for 43 minutes or 54 $\tau_{\rm A}$ (using 
the average Alfv\'{e}n time scale $\tau_{\rm A}=$48 s) including two linear 
ramps of 14.3 min each to switch this twisting motion on and off. Then the 
system evolves in a viscous relaxation for another 43 minutes to reach a 
quasi-static state where the lower smaller loops are further sheared while the 
overlying larger loops become closer to potential field, as shown in 
Figure~\ref{finit}(b). The time evolution of the kinetic energy $E_K$ and the 
magnetic energy $E_B$ of the whole system during this phase is shown in 
Figure~\ref{fenergy}(a). In order to understand the temporal evolution of 
$E_B$, we quantify its time derivative, and compare it to the total net 
Poynting flux through the six faces of the box, and the power of the 
Lorentz force in the full domain. These correspond to the first, second, and 
third term in the following formula derived from the conservation law of 
magnetic flux as
\begin{equation}
\frac{\partial}{\partial t}\int_V \frac{B^2}{2\mu_0}dV=-\int_S\frac{1}{\mu_0}
\mathbf{B}\times(\mathbf{v}\times\mathbf{B})\cdot d\mathbf{S}-\int_V 
(\mathbf{J}\times\mathbf{B})\cdot\mathbf{v}dV
\end{equation}
where $V$ is the total volume of the 3D box with its six faces represented by
$\mathbf{S}$. The quantities of these terms versus time are plotted as the 
dotted line, the dashed line, and the solid line in Figure~\ref{fenergy}(c), 
respectively. The temporal change of $E_B$ is dominated by the Poynting flux 
caused by the horizontal flows on the bottom face, while the magnitude of the 
Lorentz force power is about 50 times smaller than the Poynting flux. The 
twisting driving flows inject kinetic energy and magnetic energy into the 
domain. After the driving stops, the twisted magnetic arcade tends to relax, 
inducing a reverse rotating motion on the bottom face which emits away magnetic 
energy via the Poynting flux. This is, as mentioned before, a consequence of 
our use of ghost cell-center prescriptions to (indirectly) impose fluxes 
through this face, rather than work with a discretization which directly 
controls the face flux. Due to viscous dissipation, the kinetic energy drops 
quickly when the driving flows start to decrease and in the end relaxes to a 
very small value with a maximal speed remnant of 3.8 km s$^{-1}$. The ratio 
of the electric current that is perpendicular to the magnetic field over the 
total electric current in a volume $\mathcal{V}$:
\begin{equation}
 \sigma_\mathbf{J}\equiv=\frac{\int_\mathcal{V} |\mathbf{J}_\perp| d\mathcal{V}}
{\int_\mathcal{V} |\mathbf{J}| d\mathcal{V}}=\frac{\int_\mathcal{V} |\mathbf{J}
\times\mathbf{B}|/|\mathbf{B}| d\mathcal{V}}{\int_\mathcal{V} |\mathbf{J}| 
d\mathcal{V}}
\end{equation}
is a common measure of the degree of force-freeness of magnetic field in the 
volume $\mathcal{V}$ \citep{Valori2012SoPh}. At the end of this phase, 
$\sigma_\mathbf{J}$ is 0.055 while at the initial state it is 0.021. For a 
magnetic field that is ideally force-free, $\sigma_\mathbf{J}$ equals zero.

\subsection{Converging flows prescription}
From the end state of the first phase, we start the second phase by imposing a 
converging velocity field toward the PIL ($y=0$) on the bottom boundary filling
bottom ghost cell layers with the horizontal velocity formulated as
\begin{equation}
 v_x^b=-f(t) C_1 \frac{\partial |B_m|}{\partial x} \exp(-y^2/y_d^2),~~~~
 v_y^b=-f(t) C_1 \frac{\partial |B_m|}{\partial y} \exp(-y^2/y_d^2),
\end{equation}
where $y_d=50$ Mm quantifies an additional Gaussian width parameter away from 
the PIL, $f(t)$ is a linear ramp function as in equation~\eqref{eq:6}, and the 
amplitude factor $C_1$ is chosen so that the driven speed has a maximum value
of 12.3 km s$^{-1}$ and the maximum initial Alfv\'{e}n Mach number is 0.0216.
The pattern of the converging flows is time independent, since $f(t)$ contains 
the only time-dependent part. Net converging flows 
towards the PIL in weak field regions are an effective result of magnetic 
elements diffusion caused by random supergranular motions. These tend to 
transport vertical magnetic flux from regions of high flux density to regions 
of low flux density \citep{Zirker1997SoPh,Leighton1964ApJ}. 
Figure~\ref{fdriv}(b) shows this imposed velocity field by arrows on top of the 
magnetogram $B_m$. We then once more simulate the isothermal MHD equations 
forward in time. During this phase, we keep $\mathbf{v}=\mathbf{0}$ at the four 
side boundaries, and use a limited open boundary condition at the top and 
bottom boundaries to allow both upward and downward flows to pass through the 
boundaries. This is achieved as follows. First, we extrapolate the normal 
velocity ensuring zero normal gradient across the boundary by one-sided 
3rd-order finite differences. Then, the normal velocities on the bottom and 
the top boundary are clipped to never exceed an upper limit of 10\% of the 
local Alfv\'{e}n speed to prevent destabilisation by possible high speed 
flows near boundaries. The speed of pure downflows at the bottom is further 
limited to not be faster than 1\% of the local Alfv\'en speed. Hence at the 
bottom near the PIL, induced vertical flows are allowed, but will 
preferentially go upwards. Note that these induced flows can match or even 
exceed the imposed horizontal driving velocity magnitude. The density on the 
bottom boundary is extrapolated by gravitational stratification and kept no less
 than its initial value. The density on the other boundaries and the magnetic 
field prescriptions on all boundaries are set up in the same way as in the first
 phase. The converging motion drives the system for 50 minutes including 
linear on-off ramping phases. Note that the velocity on the bottom face is not
imposed to be zero after the driving flow stops. The time evolution of the 
kinetic energy and the magnetic energy of the whole system during this phase is 
plotted in Figure~\ref{fenergy}(b). Similar to phase 1, we plot also the time
derivative of $E_B$, the Poynting flux through the six faces of the 
box, and the power delivered by the Lorentz force versus time during this phase 
in Figure~\ref{fenergy}(d). In the first 20 min, the Poynting flux induced by 
the converging flow dominates and carries away magnetic energy through the
bottom. Later, the reconnection-induced upflows, lifting strong horizontal field
 near the PIL, dominates the Poynting flux which injects magnetic energy into
the box. After the horizontal driving flow stops, weak upflows along the PIL 
keep pumping magnetic energy into the box, while the system slowly relaxes to a 
more stable state. In the end, the maximal speed remnant is 33 km s$^{-1}$ near 
the top boundary. It is during this phase that we witness the formation of a 
large-scale flux rope structure, and its eventual relaxation to a stable 
configuration as described in the following sections.

\section{RESULTS}\label{resu}

\subsection{Birth of A Flux Rope}\label{frbirth}

\begin{figure}
\includegraphics[width=4.5in]{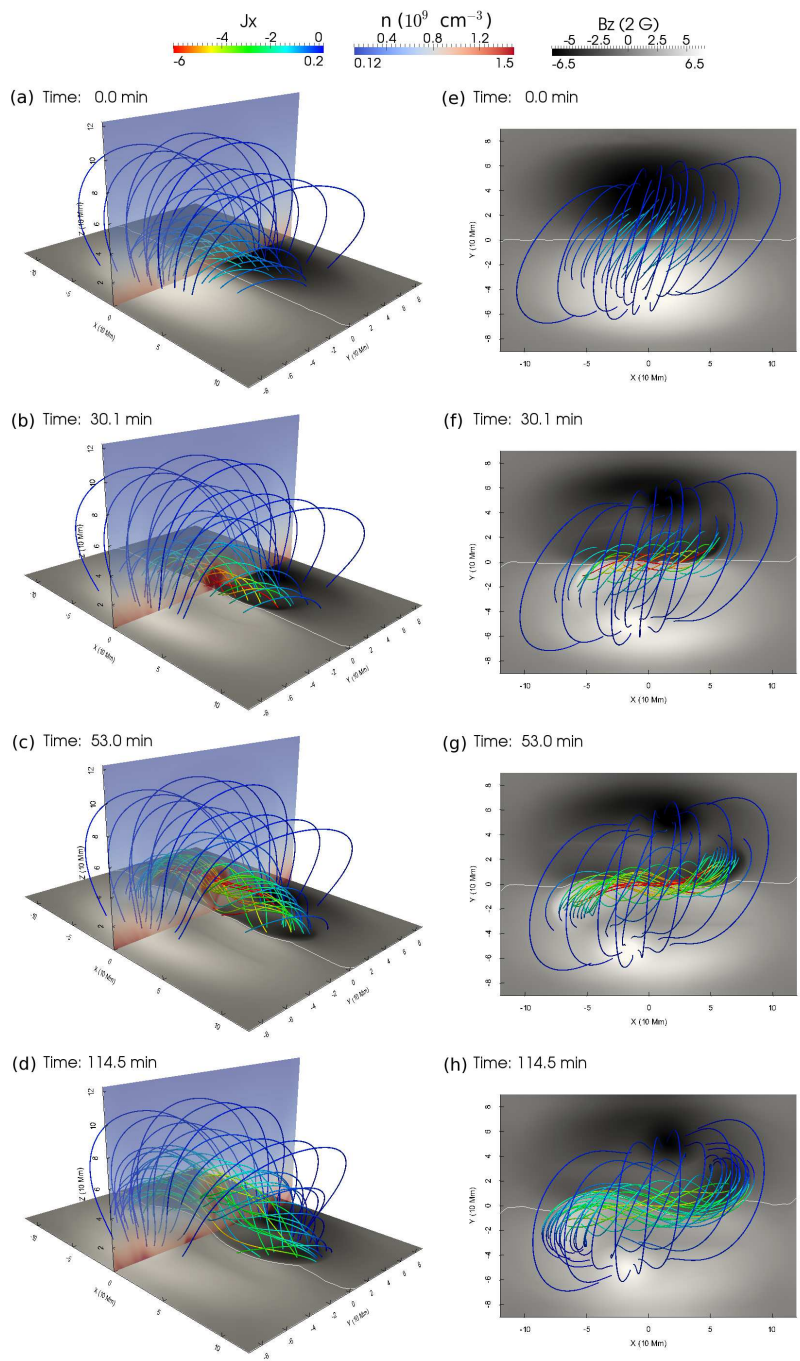}
\caption{Snapshots of the flux rope formation process at time 0 (first row),
30.1 (second row), 53 (third row), and 114.5 (bottom row) minutes. The bottom 
magnetograms are shown in gray with the PIL plotted in white. Magnetic field
lines are colored by $x$-component of the local current density $J_x$ in the 
rainbow color table. The translucent vertical planes are colored by number 
density in the blue-red color table. Side views and top views are shown in the 
left and right column respectively.
(A color version of this figure is available in the online journal.)
}
\label{fevolve}
\end{figure}

As the converging flows in frozen-in conditions force the footpoints of magnetic
 loops to approach the central PIL, the loops get sheared even further. This is 
seen in Figure~\ref{fevolve}(a)--(b). By construction, the loops most 
efficiently affected by the converging flows are rooted in the regions closer to
 the PIL, while loops rooted near the polarity centers or further away from the 
PIL are roughly unchanged, as can be seen in Figure~\ref{fevolve}(e)--(h). In 
the regions near the PIL, more and more magnetic flux of the loops is 
transferred to the direction parallel to the PIL. This is consistent with 
acquired knowledge from observations, stating that before the appearance of a 
filament near active regions, H$\alpha$ fibril structures (likely tracing the 
local magnetic field lines) change their orientation from nearly perpendicular 
to nearly parallel to the PIL \citep{Wang2007ApJ}. As the footpoints rooted in 
flux elements of opposite polarities are forced to collide at the PIL, magnetic 
reconnection happens there. As a result of the magnetic reconnection near their 
footpoints, pairs of arched magnetic flux tubes originally separate in 
$x$-extension now become linked together in a head-tail style, producing a long 
helical flux tube represented by the red central field line in Figure~
\ref{fevolve}(b)(f). It is to be noted that our ideal MHD run leaves the 
reconnection process to the inherent numerical diffusion in the employed 
numerical discretization. We will however provide evidence for the physical 
correctness of its manifestation in the next section. This newborn helical flux
 tube has a centrally concave magnetic dip, and hence ascends due to magnetic 
tension, while the magnetic reconnection below it continues to generate new 
helical flux tubes. These new helical flux tubes form and wrap around 
prior-formed ones and together they assemble into a large scale helical flux 
rope, which rises, expands, and stretches overlying loops. 

\begin{figure}
\includegraphics[width=5.5in]{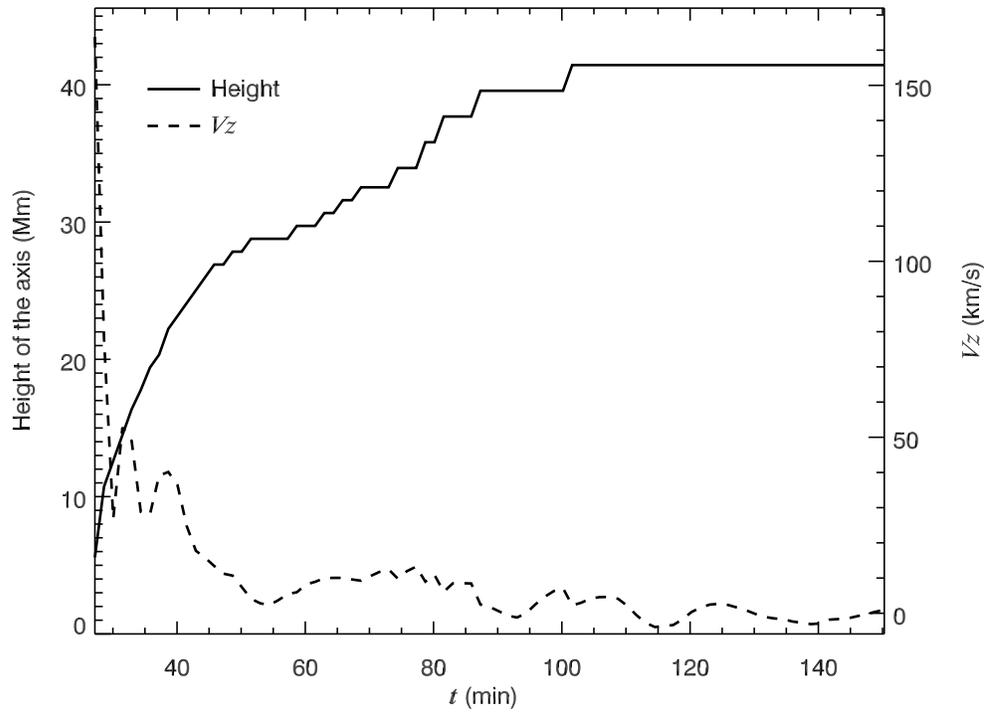}
\caption{Time evolution of the height of the axis of the flux rope (solid 
line) and the remnant local vertical velocity $V_z$ of plasma (dashed line).}
\label{faxis}
\end{figure}

A translucent vertical slice ($oyz$ plane), perpendicular to the $x$-axis, cuts 
through the center of the box in each panel of Figure~\ref{fevolve}(a)--(d) and 
it is colored by the local density. We define the axis of the flux rope by a 
central magnetic field line on which the poloidal magnetic field vanishes.
Since the flux rope perpendicularly goes through the vertical slice, the 
poloidal magnetic field in the plane can be approximately quantified by 
$B_{pol}=\sqrt{B_y^2+B_z^2}$. The axis of the flux rope intersects the plane at
the point where $B_{pol}$ takes the smallest value. In practice, the
position of this point is approximately represented by the midpoint position of
 the grid cell where $B_{pol}$ reaches the smallest value in the plane. We plot
 the height of this point and the local vertical plasma velocity there versus 
time to represent the evolution of the axis in Figure~\ref{faxis}. Note 
that the plot starts from time $t=$28.6 min, since before this time the axis of
 the flux rope can hardly be identified with sufficient resolution. From time 
28.6 minutes until the driving flows cease at time 50 min, the average 
ascending speed of the axis of the flux rope is about 13.3 km s$^{-1}$ which is
 close to the bottom driving speed. Since the ascending magnetic dips also 
carry along denser plasma upwards from lower altitude, the density inside the 
flux rope structure ends up to be higher than its surroundings, as shown by the
 density distribution in the translucent vertical slices in Figure 
\ref{fevolve}(b)--(d). After ceasing the converging driving flows at time 50 
min, the flux rope first still continues to rise and expand (see Figure 
\ref{fevolve}(c)--(d)\&(g)--(h) and Figure~\ref{faxis}). The strong current 
along the axis of the flux rope is attenuated in this expansion phase. At the 
end of the run at time 114.5 min, the system reaches a stable state, as will 
be demonstrated further on by quantifying the force balance. Moreover, 
when we continue the run to time 150 min, the height of the flux rope axis 
does not change anymore beyond time 103 min and the local speed on axis is 
nearly zero (see Figure~\ref{faxis}, the remnant velocities cause minor 
displacements that remain within the local grid cell further on). The vertical
 cross section of this mature flux rope is roughly elliptic rather than 
circular, as shown by the vertical slice of
density in Figure~\ref{fevolve}(d) and more clear in the slice of $x$-component
current density in Figure~\ref{fbala}(a). The vertical extension of the central
part of the flux rope ranges from 14 Mm to 60 Mm, and the horizontal extension
perpendicular to its axis is about 40 Mm. The ratio of the horizontal extension 
over the vertical one in the flux rope central cross section is 0.87. This 
elliptic shape of the flux rope cross-section is consistent with the elliptic 
shape often observed of coronal cavities \citep{Forland2011AGU}, and aided 
by incorporating gravity which was ignored in earlier zero-$\beta$ flux rope
 formation studies \citep{Amari1999ApJ}. To estimate the volume and the mass 
that the flux rope contains, we empirically find that the regions where 
the absolute value of $B_x J_x$, i.e. the product of the $x$-component 
magnetic field and of current, goes over a threshold of $3.18\times10^{-10}$ 
dyne cm$^{-3}$ roughly fill the volume of the flux rope. The 
estimated volume and the mass is in the ranges of $2.3$--$3.2\times10^{29}$ 
cm$^{3}$ and 4--$5\times10^{14}$ g respectively. The mass of a prominence 
estimated by observations \citep{Gilbert2011ApJ} is in the range of 
$1.04\times10^{13}$--$2.14\times10^{14}$ g which is less than the mass of our 
flux rope. Roughly half the volume of our flux rope is threaded by dipped 
field lines that can gather and stably support prominence mass in their
magnetic dips, so the mass in those dipped field lines is still significant 
compared to the mass of a prominence. Moreover, it has been shown that the 
thermal instability can induce catastrophic cooling to form prominence 
condensations along field lines \citep{Xia11, Xia2012ApJ}, but this requires 
full energetic processes to be simulated.

\subsection{Details of Magnetic Reconnection}\label{prec}

\begin{figure}
\includegraphics[width=6.2in]{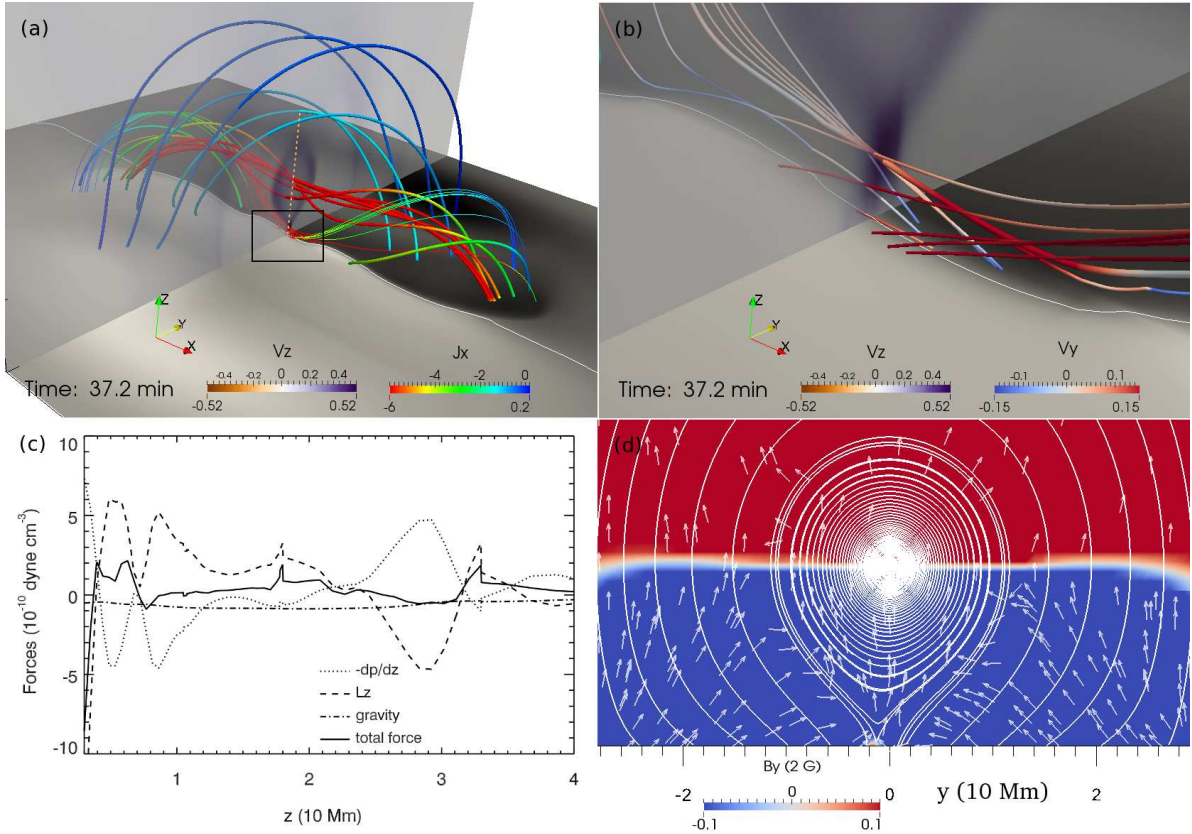}
\caption{Magnetic reconnection near the PIL at time 37.2 minutes.
(a) Overview of the large-scale flux rope colored by 
$J_x$ in rainbow color table with vertical translucent plane colored by vertical
 velocity $V_z$ in brown-purple color table; (b) Zoom-in view of the 
reconnection site delimited by the black rectangle in (a), where the field lines
are now colored by $V_y$ in a blue-red color table; (c) Distributions of 
vertical forces, such as Lorentz force L$_z$ (dashed line), pressure gradient 
force (dotted line), gravity (dash dotted line), and the total force (solid 
line), along the vertical yellow dotted line in (a); (d) Poloidal 
magnetic field lines and arrows of flow directions on top of the vertical 
plane $oyz$ which is colored by $y$-component magnetic field saturated between 
$\pm0.2$ Gauss. Velocity values are dimensionless with a unit of 116.45 km 
s$^{-1}$.
(A color version of this figure is available in the online journal.) }
\label{frecon}
\end{figure}

To find evidence in favor of magnetic reconnection, we investigate the system
 in detail at time 37.2 min when the converging flows are still ongoing. As 
shown in Figure~\ref{frecon}(a), the growing flux rope and its overarching 
arcade are clearly identifiable and indicated by representative field 
lines colored by $J_x$ in a rainbow color table. The vertical 
translucent plane in this figure where vertical velocities are quantified is 
cutting through the flux rope, and shows that the highest ascending speed 
regions locate in the outer layers of the flux rope and near but above the 
bottom surface along the PIL. We also plot some field lines threading through 
a site of magnetic reconnection at the PIL. To show these field lines clearly, 
a zoom-in view of the region in the black rectangle in Figure~\ref{frecon}(a) 
is plotted at right in Figure~\ref{frecon}(b). There, the field lines are 
colored by the $y$-component of velocities $V_y$ in a blue-red color table. The 
red field line sections move in the positive $y$-direction and the blue in the 
negative $y$-direction. This allows us to identify that field lines from 
different polarities are approaching each other, leading to magnetic 
reconnection at intersections near the PIL. These field lines are clearly driven
 towards each other by flows speeds of 18--23 km s$^{-1}$, which exceed the 
local driving speed (about 12 km s$^{-1}$) at the bottom surface. The end result
 causes topological changes in the field line patterns fully consistent with 
ongoing magnetic reconnection. In Figure~\ref{frecon}(b), the thick field line
represents a newly reconnected helical magnetic field line that joins 
in the flux rope as its outermost layer. We cut a vertical slice through the 
central part of the flux rope in Figure~\ref{frecon}(d) which shows the 
poloidal magnetic field lines, velocity arrows, and $y$-component of magnetic 
field in blue-red colors saturated at $\pm 0.2$ Gauss. The concentric ellipses
of poloidal field line show the cross section of the flux rope where the outer
layer has a tear-drop shape with a cusp reaching down close to the bottom.
Under the cusp, the $y$-component magnetic field changes sign indicating a 
small arcade and a X-point between them. The axis of the flux rope goes 
through the center of the concentric ellipses, at which height the y-component 
magnetic field changes sign. The randomly located arrows show directions of
 the local velocity in the plane. The interpretation is that as the flux rope 
is rising, the plasma at lower corona outside the flux rope is moving toward it
 driven by the bottom converging motion and sucked into the X-point. To 
understand the dynamics, we plot in Figure~\ref{frecon}(c) the vertical force 
distribution at this time along the yellow dotted line in Figure~
\ref{frecon}(a). The Lorentz force and pressure gradient force oppose each 
other and vary significantly at different heights. Below 3.7 Mm a strong 
downward Lorentz force dominates due to the magnetic tension in the concave 
down small loops. Above a height of 4 Mm up to 25 Mm inside the flux rope, the 
Lorentz force is upward, as well as the total force, which leads to creating 
newborn flux tubes and also their ascending in the flux rope as shown by the 
vertical velocity distribution in purple in the vertical plane. The strongest 
upflows (up to 60 km s$^{-1}$) appear just above the sites of magnetic 
reconnection. The gravity is relatively small and dynamically only enhanced a 
little inside the flux rope due to the density enhancement there. From 25 Mm 
to 31.5 Mm the Lorentz force is downward again, which means that the ascending 
flux rope is ultimately restrained by the overlying magnetic arcade. Since
 we are limited to an isothermal model, the thermal energy release of magnetic 
reconnection is not captured, but the conservative nature of the numerical 
scheme ensures overall mass, momentum and magnetic flux evolutions consistent 
with the (partially open) boundary prescriptions, while the energy evolution 
is constrained to yield an isothermal endstate.

\subsection{Overall Force Balance}\label{force}

\begin{figure}
\includegraphics[width=6.2in]{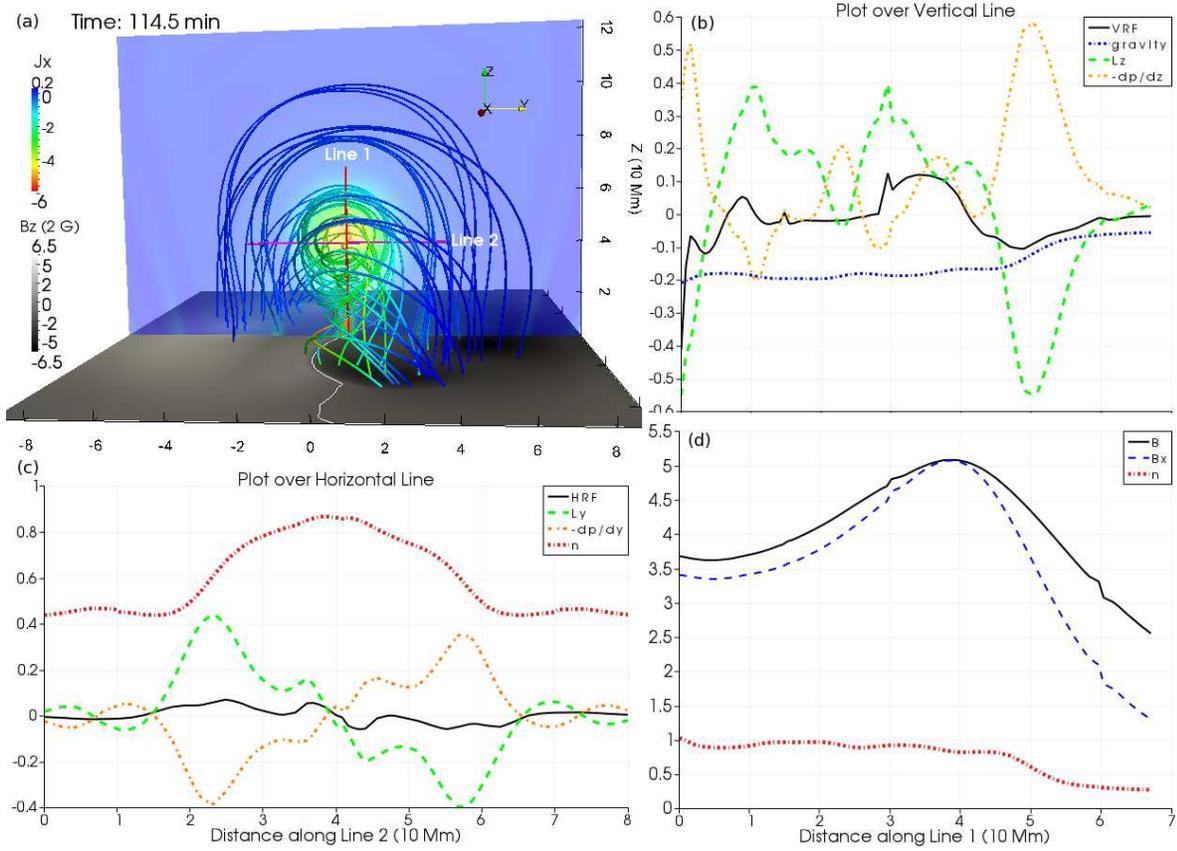}
\caption{Analysis of the end state at time 114.5 min. (a) a view through the 
axis of the flux rope with field lines as well as the vertical cutting 
plane colored by $J_x$; (b) distributions of gravity (blue dotted line), 
vertical Lorentz force (green dashed line), vertical pressure gradient force 
(orange dotted line), and vertical resultant force (black line) along Line 1 
shown in panel (a); (c) distributions of number density (red dotted line), 
horizontal Lorentz force (green dashed line), horizontal pressure gradient 
force (orange dotted line), and horizontal resultant force (black line) along 
Line 2 in panel (a); (d) distributions of magnetic field strength (black 
line), $x$-component of magnetic field (blue dashed line), and number density 
(red dotted line) along Line 1 shown in panel (a). The line plot 
vertical ranges use dimensionless values with these units: $3.2\times
10^{-10}$ dyne cm$^{-3}$ for forces, $10^9$ cm$^{-3}$ for number density, and
 2 Gauss for magnetic field.
(A color version of this figure is available in the online journal.)
}
\label{fbala}
\end{figure}

To understand the stability of the flux rope at the end of the simulation, we
quantify the forces along two lines, vertical Line 1 and horizontal Line 2,
which go across the central part of the flux rope as shown in
Figure~\ref{fbala}(a). In the vertical direction, the gravity, Lorentz force,
the pressure gradient force, and the vertical resultant force (VRF) are
quantified along Line 1 shown by Figure~\ref{fbala}(b). The Lorentz force and
the pressure gradient force fluctuate and oppose each other. Combined with 
 gravity, the VRF is roughly around zero, which means the flux rope is overall
stabilized vertically. Strong negative (downward) Lorentz force and the VRF near
 the very bottom are results of the magnetic striction of low-lying small 
arcades under the flux rope. In the upper layer of the flux rope, downward 
Lorentz force tends to hold down the inflating flux rope. In the rest part, the
 positive (upward) Lorentz force shows a lifting tend and is compensated by the
 gravity and gas pressure effect. The distributions of total magnetic field 
strength, $x$-component magnetic field, and the number density are shown in 
Figure~\ref{fbala}(d). The number density deviates from the hydrostatic state 
with an enhancement inside the flux rope, and so does the influence of gravity.
 The magnetic field strength is enhanced inside the flux rope and reaches its 
maximum at the axis of the flux rope, where the poloidal magnetic field is zero
 ($B=B_x$). In the horizontal direction along Line 2, the Lorentz force in the 
flux rope, which points to the center from two sides , is almost compensated 
by the gas pressure gradient force pointing outward from the center. The 
horizontal resultant force (HRF) is nearly zero showing a force balance in the 
horizontal direction. The number density inside the flux rope is enhanced and 
reaches the maximum (2 times denser than the same altitude outside the flux 
rope) at the center. The flux rope is self pinching and resisted by gas 
pressure. Thus the enhancement of the density inside the flux rope has two 
reasons: (1) higher density plasma from lower atmosphere being brought up, (2) 
squeezing effect by the Lorentz force on the flux rope. The degree of 
force-freeness of the magnetic field in the computational box now is 0.086, 
increased by 56\% from the beginning state of the second phase. The magnetic 
field of the system deviates further away from force-free field.

\subsection{Twist of the Flux Rope}\label{twist}
\begin{figure}
\includegraphics[width=5.6in]{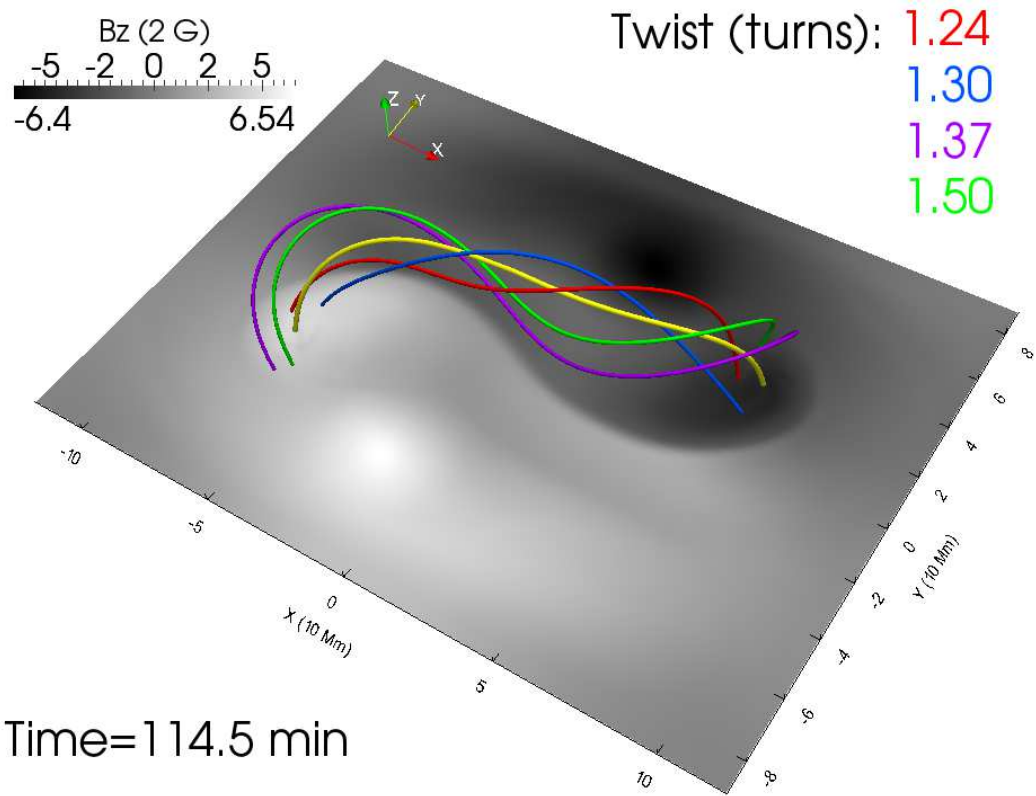}
\caption{The twists (in turns) of four field lines of the flux rope winding 
around the axis field line (yellow line) are 1.24 (red line), 1.30 (blue line), 
1.37 (purple line), and 1.50 (green line) at the time 114.5 min.
(A color version of this figure is available in the online journal.)
}
\label{ftwist}
\end{figure}

To quantify the helical property of the flux rope and to check whether the kink
instability might happen, we compute the twist of the flux rope. The twist
density (in the unit of turns) of a curve $\mathbf{B}$ around a smooth axis 
curve $\mathbf{A}$ is defined as 
\begin{equation}
\frac{d \phi}{d s}=\frac{1}{2 \pi}
\mathbf{T}(s)\cdot\left(\mathbf{V}(s)\times\frac{d\mathbf{V}(s)}{d s}\right)
\end{equation}
where $s$ is the distance along the axis curve, $\mathbf{T}(s)$ is a tangential
unit vector of $\mathbf{A}$, and $\mathbf{V}(s)$ is a unit vector normal to 
$\mathbf{T}(s)$ and pointing from the point $\mathbf{A}(s)$ to a point on 
$\mathbf{B}$ \citep{Guo2010ApJ,Berger2006JPhA}. The integration of this equation
 along the axis curve $\mathbf{A}$ gives the twist of the curve $\mathbf{B}$. 
We select few representative magnetic field lines that wind the axis of the 
flux rope in the flux rope, and compute their twists. The result is shown in 
Figure~\ref{ftwist}. The average twist 1.35 is smaller than the lower limits of 
the twist for kink-unstable coronal flux ropes of 1.5 and 1.75 turns found by
\citet{Fan2003ApJ} and \citet{Torok2004AA}, respectively. So our flux rope is
kink-stable. In the middle of the flux rope, the field lines that go above 
the axis have smaller twist than the field lines that pass it from below.

\section{CONCLUSIONS AND DISCUSSION}\label{conclusion}

We reported on the self-consistent formation of a 3D stable, large-scale, 
plasma carrying flux rope by successive footpoint shearing and converging flows,
followed by relaxations towards near-perfect force-balanced endstates. We 
quantitatively analyzed the evolution of the magnetic energy, the degree of 
non force-freeness in the flux rope, the accumulated mass, and its most 
important geometric features. Improvements to previous efforts are  
the combination of (1) the gravitationally stratified atmosphere incorporated 
at finite $\beta$; (2) the elliptic cross-sectional flux rope shape in 
agreement with recent observations; (3) the equilibrium balance between pressure
 gradient, Lorentz force and gravity as demonstrated in the endstate; and (4) 
the fact that a kink-stable flux rope persists on a long timescale, itself 
trapping sufficient plasma to potentially condense into a true 
prominence within the flux rope. The latter aspect can only be demonstrated 
fully when relaxing the assumption of isothermal conditions assumed thus far. By
 analyzing the numerical evolution scenario, we presented convincing arguments 
for the role of the induced reconnection near and just above the PIL, with 
consistent flow and magnetic rearrangements gradually building up a large-scale 
flux rope. By showing the energetic evolution, as well as the position of the 
magnetic axis, it is clear that the overall configuration is dominated by 
magnetic energy stored in the helical flux rope, itself mostly held down by the 
Lorentz force of the overlying arcade field. This endstate can now be used for 
systematic studies of realistic (plasma-carrying) coronal mass ejections, by 
further energizing the flux rope structure by e.g. twisting motions driving the 
twist above kink-unstable values, although that may require even larger 
simulation boxes than incorporated here. We also plan to extend this work to 
fully thermodynamically consistent 3D scenarios. This can be done along the 
lines already demonstrated for 2.5D magnetic arcades where filaments form due 
to chromospheric evaporation and thermal instability, on top of the arcade 
\citep{Xia2012ApJ}.

\acknowledgments
This research was supported by projects GOA/2009/009 (KU Leuven), the EC 
seventh framework programme (FP7/2007-2013) under grant agreement Swiff 
(proj. no. 263340) and by the Interuniversity Attraction Poles Programme 
initiated by the Belgian Science Policy Office (IAP P7/08 CHARM). Part of the 
simulations used the VSC (flemish supercomputer center) funded by the Hercules 
foundation and the Flemish government. Part of the results were obtained using 
the Curie supercomputer made available by the PRACE allocation number 
2011050747. We acknowledge fruitful discussions with Jie Zhao, Xia Fang, and 
Gherardo Valori and we thank the referee for revisions and complementary 
opinions. YG was supported by the National Natural Science Foundation of
China (NSFC) under the grant numbers 11203014, 10933003, and the grant from the
 973 project 2011CB811402.


\begin{thebibliography}{54}
\expandafter\ifx\csname natexlab\endcsname\relax\def\natexlab#1{#1}\fi

\bibitem[{{Amari} {et~al.}(2010){Amari}, {Aly}, {Mikic}, \& {Linker}}]{Amari10}
{Amari}, T., {Aly}, J.-J., {Mikic}, Z., \& {Linker}, J. 2010, \apjl, 717, L26

\bibitem[{{Amari} {et~al.}(2003{\natexlab{a}}){Amari}, {Luciani}, {Aly},
  {Mikic}, \& {Linker}}]{Amari03a}
{Amari}, T., {Luciani}, J.~F., {Aly}, J.~J., {Mikic}, Z., \& {Linker}, J.
  2003{\natexlab{a}}, \apj, 585, 1073

\bibitem[{{Amari} {et~al.}(2003{\natexlab{b}}){Amari}, {Luciani}, {Aly},
  {Mikic}, \& {Linker}}]{Amari03b}
---. 2003{\natexlab{b}}, \apj, 595, 1231

\bibitem[{{Amari} {et~al.}(1996){Amari}, {Luciani}, {Aly}, \&
  {Tagger}}]{Amari96}
{Amari}, T., {Luciani}, J.~F., {Aly}, J.~J., \& {Tagger}, M. 1996, \apjl, 466,
  L39

\bibitem[{{Amari} {et~al.}(1999){Amari}, {Luciani}, {Mikic}, \&
  {Linker}}]{Amari1999ApJ}
{Amari}, T., {Luciani}, J.~F., {Mikic}, Z., \& {Linker}, J. 1999, \apjl, 518,
  L57

\bibitem[{{Amari} {et~al.}(2000){Amari}, {Luciani}, {Mikic}, \&
  {Linker}}]{Amari2000}
---. 2000, \apjl, 529, L49

\bibitem[{{Bak-St{\c e}{\'s}licka} {et~al.}(2013){Bak-St{\c e}{\'s}licka},
  {Gibson}, {Fan}, {Bethge}, {Forland}, \& {Rachmeler}}]{Bak2013ApJ}
{Bak-St{\c e}{\'s}licka}, U., {Gibson}, S.~E., {Fan}, Y., {Bethge}, C.,
  {Forland}, B., \& {Rachmeler}, L.~A. 2013, \apjl, 770, L28

\bibitem[{{Berger} \& {Prior}(2006)}]{Berger2006JPhA}
{Berger}, M.~A. \& {Prior}, C. 2006, Journal of Physics A Mathematical General,
  39, 8321

\bibitem[{{Berger}(2012)}]{Berger2012ASPC}
{Berger}, T. 2012, in Astronomical Society of the Pacific Conference Series,
  Vol. 463, Astronomical Society of the Pacific Conference Series, ed. T.~R.
  {Rimmele}, A.~{Tritschler}, F.~{W{\"o}ger}, M.~{Collados Vera},
  H.~{Socas-Navarro}, R.~{Schlichenmaier}, M.~{Carlsson}, T.~{Berger},
  A.~{Cadavid}, P.~R. {Gilbert}, P.~R. {Goode}, \& M.~{Kn{\"o}lker}, 147

\bibitem[{{Blokland} \& {Keppens}(2011)}]{Blok2011AA}
{Blokland}, J.~W.~S. \& {Keppens}, R. 2011, \aap, 532, A93

\bibitem[{{Burlaga} {et~al.}(1982){Burlaga}, {Klein}, {Sheeley}, {Michels},
  {Howard}, {Koomen}, {Schwenn}, \& {Rosenbauer}}]{Burlaga1982GeoRL}
{Burlaga}, L.~F., {Klein}, L., {Sheeley}, Jr., N.~R., {Michels}, D.~J.,
  {Howard}, R.~A., {Koomen}, M.~J., {Schwenn}, R., \& {Rosenbauer}, H. 1982,
  \grl, 9, 1317

\bibitem[{{Chae} {et~al.}(2001){Chae}, {Wang}, {Qiu}, {Goode}, {Strous}, \&
  {Yun}}]{Chae2001ApJ}
{Chae}, J., {Wang}, H., {Qiu}, J., {Goode}, P.~R., {Strous}, L., \& {Yun},
  H.~S. 2001, \apj, 560, 476

\bibitem[{{Chiu} \& {Hilton}(1977)}]{Chiu1977ApJ}
{Chiu}, Y.~T. \& {Hilton}, H.~H. 1977, \apj, 212, 873

\bibitem[{{DeVore} \& {Antiochos}(2000)}]{DeVor00}
{DeVore}, C.~R. \& {Antiochos}, S.~K. 2000, \apj, 539, 954

\bibitem[{{DeVore} {et~al.}(2005){DeVore}, {Antiochos}, \&
  {Aulanier}}]{DeVor05}
{DeVore}, C.~R., {Antiochos}, S.~K., \& {Aulanier}, G. 2005, \apj, 629, 1122

\bibitem[{{Fan}(2001)}]{Fan01}
{Fan}, Y. 2001, \apjl, 554, L111

\bibitem[{{Fan}(2010)}]{Fan2010ApJ}
---. 2010, \apj, 719, 728

\bibitem[{{Fan} \& {Gibson}(2003)}]{Fan2003ApJ}
{Fan}, Y. \& {Gibson}, S.~E. 2003, \apjl, 589, L105

\bibitem[{{Forland} {et~al.}(2011){Forland}, {Rachmeler}, {Gibson}, \&
  {Dove}}]{Forland2011AGU}
{Forland}, B., {Rachmeler}, L.~A., {Gibson}, S.~E., \& {Dove}, J. 2011, AGU
  Fall Meeting Abstracts, B1951

\bibitem[{{Foukal}(1971)}]{Foukal1971SoPh}
{Foukal}, P. 1971, \solphys, 19, 59

\bibitem[{{Gibson} {et~al.}(2010){Gibson}, {Kucera}, {Rastawicki}, {Dove}, {de
  Toma}, {Hao}, {Hill}, {Hudson}, {Marqu{\'e}}, {McIntosh}, {Rachmeler},
  {Reeves}, {Schmieder}, {Schmit}, {Seaton}, {Sterling}, {Tripathi},
  {Williams}, \& {Zhang}}]{Gibson2010ApJ}
{Gibson}, S.~E., {Kucera}, T.~A., {Rastawicki}, D., {Dove}, J., {de Toma}, G.,
  {Hao}, J., {Hill}, S., {Hudson}, H.~S., {Marqu{\'e}}, C., {McIntosh}, P.~S.,
  {Rachmeler}, L., {Reeves}, K.~K., {Schmieder}, B., {Schmit}, D.~J., {Seaton},
  D.~B., {Sterling}, A.~C., {Tripathi}, D., {Williams}, D.~R., \& {Zhang}, M.
  2010, \apj, 724, 1133

\bibitem[{{Gilbert} {et~al.}(2011){Gilbert}, {Kilper}, {Alexander}, \&
  {Kucera}}]{Gilbert2011ApJ}
{Gilbert}, H., {Kilper}, G., {Alexander}, D., \& {Kucera}, T. 2011, \apj, 727,
  25

\bibitem[{{Guo} {et~al.}(2010){Guo}, {Ding}, {Schmieder}, {Li},
  {T{\"o}r{\"o}k}, \& {Wiegelmann}}]{Guo2010ApJ}
{Guo}, Y., {Ding}, M.~D., {Schmieder}, B., {Li}, H., {T{\"o}r{\"o}k}, T., \&
  {Wiegelmann}, T. 2010, \apjl, 725, L38

\bibitem[{{Hillier} \& {van Ballegooijen}(2013)}]{Hillier2013ApJ}
{Hillier}, A. \& {van Ballegooijen}, A. 2013, \apj, 766, 126

\bibitem[{{Keppens} {et~al.}(2012){Keppens}, {Meliani}, {van Marle}, {Delmont},
  {Vlasis}, \& {van der Holst}}]{Kepp12}
{Keppens}, R., {Meliani}, Z., {van Marle}, A.~J., {Delmont}, P., {Vlasis}, A.,
  \& {van der Holst}, B. 2012, Journal of Computational Physics, 231, 718

\bibitem[{{Keppens} {et~al.}(2003){Keppens}, {Nool}, {T{\'o}th}, \&
  {Goedbloed}}]{Keppens03}
{Keppens}, R., {Nool}, M., {T{\'o}th}, G., \& {Goedbloed}, J.~P. 2003, Computer
  Physics Communications, 153, 317

\bibitem[{{Leighton}(1964)}]{Leighton1964ApJ}
{Leighton}, R.~B. 1964, \apj, 140, 1547

\bibitem[{{Leroy} {et~al.}(1983){Leroy}, {Bommier}, \&
  {Sahal-Brechot}}]{Leroy1983SoPh}
{Leroy}, J.~L., {Bommier}, V., \& {Sahal-Brechot}, S. 1983, \solphys, 83, 135

\bibitem[{{Linker} {et~al.}(2001){Linker}, {Lionello}, {Miki{\'c}}, \&
  {Amari}}]{Linker01}
{Linker}, J.~A., {Lionello}, R., {Miki{\'c}}, Z., \& {Amari}, T. 2001, \jgr,
  106, 25165

\bibitem[{{Linker} {et~al.}(2003){Linker}, {Miki{\'c}}, {Lionello}, {Riley},
  {Amari}, \& {Odstrcil}}]{Linker03}
{Linker}, J.~A., {Miki{\'c}}, Z., {Lionello}, R., {Riley}, P., {Amari}, T., \&
  {Odstrcil}, D. 2003, Physics of Plasmas, 10, 1971

\bibitem[{{Lionello} {et~al.}(2002){Lionello}, {Miki{\'c}}, {Linker}, \&
  {Amari}}]{Lione02}
{Lionello}, R., {Miki{\'c}}, Z., {Linker}, J.~A., \& {Amari}, T. 2002, \apj,
  581, 718

\bibitem[{{Litvinenko} {et~al.}(2007){Litvinenko}, {Chae}, \&
  {Park}}]{Litvinenko2007ApJ}
{Litvinenko}, Y.~E., {Chae}, J., \& {Park}, S.-Y. 2007, \apj, 662, 1302

\bibitem[{{Low} \& {Zhang}(2004)}]{Low2004ApJ}
{Low}, B.~C. \& {Zhang}, M. 2004, \apj, 609, 1098

\bibitem[{{Mackay} {et~al.}(2008){Mackay}, {Gaizauskas}, \& {Yeates}}]{Macka08}
{Mackay}, D.~H., {Gaizauskas}, V., \& {Yeates}, A.~R. 2008, \solphys, 248, 51

\bibitem[{{Mackay} \& {van Ballegooijen}(2006)}]{Mackay2006}
{Mackay}, D.~H. \& {van Ballegooijen}, A.~A. 2006, \apj, 641, 577

\bibitem[{{Martens} \& {Zwaan}(2001)}]{Marte01}
{Martens}, P.~C. \& {Zwaan}, C. 2001, \apj, 558, 872

\bibitem[{{Martin}(1998)}]{Marti98}
{Martin}, S.~F. 1998, in Astronomical Society of the Pacific Conference Series,
  Vol. 150, IAU Colloq. 167: New Perspectives on Solar Prominences, ed.
  {D.~F.~Webb, B.~Schmieder, \& D.~M.~Rust}, 419

\bibitem[{{Mignone} \& {Bodo}(2006)}]{Mignone2006MNRAS}
{Mignone}, A. \& {Bodo}, G. 2006, \mnras, 368, 1040

\bibitem[{{Okamoto} {et~al.}(2008){Okamoto}, {Tsuneta}, {Lites}, {Kubo},
  {Yokoyama}, {Berger}, {Ichimoto}, {Katsukawa}, {Nagata}, {Shibata},
  {Shimizu}, {Shine}, {Suematsu}, {Tarbell}, \& {Title}}]{Okam08}
{Okamoto}, T.~J., {Tsuneta}, S., {Lites}, B.~W., {Kubo}, M., {Yokoyama}, T.,
  {Berger}, T.~E., {Ichimoto}, K., {Katsukawa}, Y., {Nagata}, S., {Shibata},
  K., {Shimizu}, T., {Shine}, R.~A., {Suematsu}, Y., {Tarbell}, T.~D., \&
  {Title}, A.~M. 2008, \apjl, 673, L215

\bibitem[{{Savcheva} {et~al.}(2012){Savcheva}, {Pariat}, {van Ballegooijen},
  {Aulanier}, \& {DeLuca}}]{Savcheva2012}
{Savcheva}, A., {Pariat}, E., {van Ballegooijen}, A., {Aulanier}, G., \&
  {DeLuca}, E. 2012, \apj, 750, 15

\bibitem[{{T{\"o}r{\"o}k} {et~al.}(2004){T{\"o}r{\"o}k}, {Kliem}, \&
  {Titov}}]{Torok2004AA}
{T{\"o}r{\"o}k}, T., {Kliem}, B., \& {Titov}, V.~S. 2004, \aap, 413, L27

\bibitem[{{{\v C}ada} \& {Torrilhon}(2009)}]{Cada09JCoPh}
{{\v C}ada}, M. \& {Torrilhon}, M. 2009, Journal of Computational Physics, 228,
  4118

\bibitem[{{Valori} {et~al.}(2012){Valori}, {Green}, {D{\'e}moulin}, {Vargas
  Dom{\'{\i}}nguez}, {van Driel-Gesztelyi}, {Wallace}, {Baker}, \&
  {Fuhrmann}}]{Valori2012SoPh}
{Valori}, G., {Green}, L.~M., {D{\'e}moulin}, P., {Vargas Dom{\'{\i}}nguez},
  S., {van Driel-Gesztelyi}, L., {Wallace}, A., {Baker}, D., \& {Fuhrmann}, M.
  2012, \solphys, 278, 73

\bibitem[{{van Ballegooijen} \& {Martens}(1989)}]{vanBa1989ApJ}
{van Ballegooijen}, A.~A. \& {Martens}, P.~C.~H. 1989, \apj, 343, 971

\bibitem[{{van Ballegooijen} \& {Martens}(1990)}]{vanBa1990ApJ}
---. 1990, \apj, 361, 283

\bibitem[{{van der Holst} \& {Keppens}(2007)}]{vanderH07}
{van der Holst}, B. \& {Keppens}, R. 2007, Journal of Computational Physics,
  226, 925

\bibitem[{{Waldmeier}(1970)}]{Waldmeier1970SoPh}
{Waldmeier}, M. 1970, \solphys, 15, 167

\bibitem[{{Wang} \& {Muglach}(2007)}]{Wang2007ApJ}
{Wang}, Y. \& {Muglach}, K. 2007, \apj, 666, 1284

\bibitem[{{Wang} \& {Stenborg}(2010)}]{WangYM2010ApJ}
{Wang}, Y.-M. \& {Stenborg}, G. 2010, \apjl, 719, L181

\bibitem[{{Welsch} {et~al.}(2005){Welsch}, {DeVore}, \& {Antiochos}}]{Welsch05}
{Welsch}, B.~T., {DeVore}, C.~R., \& {Antiochos}, S.~K. 2005, \apj, 634, 1395

\bibitem[{{Xia} {et~al.}(2012){Xia}, {Chen}, \& {Keppens}}]{Xia2012ApJ}
{Xia}, C., {Chen}, P.~F., \& {Keppens}, R. 2012, \apjl, 748, L26

\bibitem[{{Xia} {et~al.}(2011){Xia}, {Chen}, {Keppens}, \& {van Marle}}]{Xia11}
{Xia}, C., {Chen}, P.~F., {Keppens}, R., \& {van Marle}, A.~J. 2011, \apj, 737,
  27

\bibitem[{{Yeates} \& {Mackay}(2009)}]{Yeates09}
{Yeates}, A.~R. \& {Mackay}, D.~H. 2009, \apj, 699, 1024

\bibitem[{{Zirker} {et~al.}(1997){Zirker}, {Martin}, {Harvey}, \&
  {Gaizauskas}}]{Zirker1997SoPh}
{Zirker}, J.~B., {Martin}, S.~F., {Harvey}, K., \& {Gaizauskas}, V. 1997,
  \solphys, 175, 27

\end{thebibliography}

\clearpage

\clearpage

\clearpage

\clearpage

\clearpage

\clearpage

\clearpage

\clearpage

\end{document}